\documentclass[aps,pra,twocolumn,superscriptaddress,groupedaddress]{revtex4-1}

\usepackage[pdftex]{color}
\usepackage[utf8]{inputenc}
\usepackage[english]{babel}
\usepackage{amsfonts}
\usepackage{amssymb}

\usepackage{amsmath}
\usepackage{graphicx}

\newcommand{\Ket}[1]{\left|#1\right>}

\newcommand{\BraKet}[2]{\left<#1|#2\right>}

\newcommand{\ex}[1]{\mathrm{e}^{#1}}

\begin{document}

\title{Quantum Speed Limit and optimal evolution time in a two-level system}
\author{P.M. Poggi}
\author{F.C. Lombardo}
\author{D.A. Wisniacki}
\affiliation{Departamento de F\'isica Juan Jos\'e Giambiagi FCEyN UBA and and IFIBA - CONICET,
Facultad de Ciencias Exactas y Naturales, 
Ciudad Universitaria, Pabell\'on I, 1428 Buenos Aires, Argentina.
\footnotesize Electronic Address: {\tt ppoggi@df.uba.ar}}
\date[]{December 17, 2013}


\begin{abstract}
Quantum mechanics establishes a fundamental bound for the minimum evolution time between two states of a given system. Known as the quantum speed limit (QSL), it is a useful tool in the context of quantum control, where the speed of some control protocol is usually intended to be as large as possible. While QSL expressions for time-independent Hamiltonians have been well studied, the time-dependent regime has remained somewhat unexplored, albeit being usually the relevant problem to be compared with when studying systems controlled by external fields. In this paper we explore the relation between optimal times found in quantum control and the QSL bound, in the (relevant) time-dependent regime, by discussing the ubiquitous two-level Landau-Zener type Hamiltonian.
\end{abstract}

\maketitle

\section{Introduction} 
Derivation of optimal times for the evolution of a quantum system is an essential part of the design of quantum control protocols and quantum algorithms \cite{bib:caneva,bib:chuang,bib:nos,bib:muga}. There, operations have to be performed in a rapid way to avoid undesirable environmental effects which can destroy the coherence properties of the system under consideration. The basic formulation of the time--optimal control (time--OC) problem is the following: given a quantum system and a Hamiltonian $H(u)$, the goal is to find a control function $u(t)$ such that the system, initially prepared in state $\psi_0$, evolves to a target state $\psi(\tau)=\psi_{target}$ (with probability close to 1) in the minimum possible time $\tau$ \cite{bib:dalessandro}. Usually, analytical solutions to this problem are not available, and so numerical estimations have to be drawn for each particular physical setup.

Besides its practical importance, the subject of time--OC is indeed of fundamental interest, as limits to the speed of evolution of a quantum system are imposed by the time-energy uncertainty relation, due originally to Heisenberg and later generalized by Mandelstam and Tamm \cite{bib:mt}. Along with them, Fleming \cite{bib:fleming}, Bhattacharyya \cite{bib:bhatta} and later Pfeifer \cite{bib:pfeifer} established the formulation of what is usually referred to as the Quantum Speed Limit, which states that for a quantum system subjected to a time-independent Hamiltonian $H$, initially prepared in some state $\psi_0$, the evolution time $\tau$ required to reach $\psi({\tau})$ satisfies the following inequality

\begin{equation}
\tau\geq\frac{\hbar}{\Delta E}\:\mathrm{arccos}\left(\left|\BraKet{\psi_0}{\psi(\tau)}\right|\right)\equiv\tau_{QSL},
\label{ec:mtbound}
\end{equation}

\noindent where $\Delta E^2$ is the variance of the Hamiltonian, $\Delta E^2=\langle(H-\langle H\rangle)^2\rangle$ and the expectation value can be taken either over the initial state $\Ket{\psi_0}$ or the evolutioned state $\Ket{\psi(\tau)}$, as in this conditions energy is a constant of motion. Equation (\ref{ec:mtbound}) is usually referred to as the Mandelstam-Tamm (MT) or Bhatacharyya bound. Extensions and generalizations of this problem have already been studied; for example, the quantum speed limit for open quantum systems is adressed in Refs. \cite{bib:delcampo,bib:davidovich,bib:lutz1} using different approaches; Margolus and Levitin \cite{bib:margo} proposed a bound for passage times (i.e., the special case when $\Ket{\psi_0}$ and $\Ket{\psi_{\tau}}$ are orthogonal)  which depends on the mean energy rather than on the variance of $H$, and their work was later generalized for arbitrary initial and final states \cite{bib:lloyd,bib:toffoli}. \\

When assesing quantum control protocols, the deviation of the evolution time from the QSL bound has been proposed as a natural measure \cite{bib:caneva,bib:bason,bib:heger} for the time performance of the protocol, yielding excellent performance if the QSL bound is attained by the evolution. However, the MT bound is not suitable for adressing time-dependent Hamiltonians which are, in general, the generators of the dynamics in controlled systems. Straightforward extensions of the MT relation have been put forward in the literature \cite{bib:muga,bib:anandan,bib:davidovich,bib:lutz1}. These relations are based on the concept of distance in state space of a quantum system and concur to a single inequality for the case of unitary dynamics. For time dependent systems, this approach has been proposed to lead to implicit bounds for the evolution time \cite{bib:lutz2}. \\

In this work we study the different bounds given by the usual QSL formulation for time-dependent hamiltonians, discuss their interpretation and compare their features with the well-known time-independent case. For this purpose, we analyze a paradigmatic model of a driven two-level system, for which time-optimal control problem has been analytically solved \cite{bib:heger}. In our analysis, we compare the optimal evolution times with the QSL bounds and discuss at what extent those bounds are useful for assesing the time performance of a control protocol. We show that, in some cases, no meaningful bound can be obtained even if precise knowledge of the whole physical evolution is available.


\section{QSL for time-dependent Hamiltonians and optimal quantum control} 

In order to explore in what way the QSL formulation allows us to derive bounds for evolution times in quantum control problems, we will begin by revisiting the generalization of the MT bound. Consider a generic quantum system subjected to a time-dependent Hamiltonian $H(t)$. The original derivation of the MT bound shows that the following relation is always satisfied

\begin{equation}
2\:\mathrm{arccos}\left(\left|\BraKet{\psi_0}{\psi(\tau)}\right|\right)\leq 2\int_0^\tau\: \Delta E (t')\:dt'.
\label{ec:cota}
\end{equation} 

In the following we will restrict ourselves to unitary dynamics and set $\hbar=1$. It is straightforward to see that when $H\neq H(t)$, then $\Delta E=\mathrm{const.}$ and after solving the integral in expression (\ref{ec:cota}) we recover the MT bound, eq. (\ref{ec:mtbound}). As has already been noted in previous works \cite{bib:anandan,bib:brody1,bib:geom}, the inequality (\ref{ec:cota}) is geometric in nature, as it states the fact that the distance between two states (l.h.s.), as measured by the Fubini-Study metric

\begin{equation}
s(\psi,\phi)=2\:\mathrm{arccos}\left(\left|\BraKet{\psi}{\phi}\right|\right),
\label{ec:dist}
\end{equation}

\noindent is always smaller than or equal to the length of the actual path followed by the evolution in state space. This interpreation is due to Anandan and Aharonov \cite{bib:anandan} who showed that the quantity on the l.h.s. of eq. (\ref{ec:cota}) is independent of the actual Hamiltonian used to generate the evolution and is thus a purely geometric quantity. Moreover, they demonstrated that the speed of the system in space state is given by $\frac{ds}{dt}=2\:\Delta E(t)$. Note that $\Delta E(t)$ here has to be calculated over the evolutioned state, as now energy is not constant during the evolution. This means that, in general, the complete solution for the evolution operator $U(t)$ has to be known in order to evaluate the r.h.s. of expression (\ref{ec:cota}). We remark that this relation can be obtained as a special case of a more general bound in terms of the quantum Fisher information \cite{bib:brody2,bib:davidovich}. \\

Note that the equality in eq. (\ref{ec:cota}) holds if and only if the evolution of the system takes place following the shortest path between the initial and final states, that is, following a geodesic. This solution usually corresponds to the ``Quantum Brachistochrone'' problem \cite{bib:carlini,bib:brody3}, where the goal is to find the Hamiltonian which connects two different states in the minimum possible time, given a set of dynamical constraints. We remark the subtle difference between this problem and time-OC, where constrains are imposed  though specifying the structure of the Hamiltonian $H(u)$ and optimization is achieved through the determination of the control field $u(t)$. In this case, if $H(u)$ is incompatible with the generator of the geodesic path, then there will be no process for which the equality in expression (\ref{ec:cota}) holds. \\

We now turn to the problem of bounding evolution times in a quantum control scenario. Suppose an specific control field $u(t)$ is given such that $H(t)$ connects $\Ket{\psi_0}$ and $\Ket{\psi_g}$ in a time $T$. In order to obtain a QSL time for this process from relation (\ref{ec:cota}), we can set $\Ket{\psi(\tau)}=\Ket{\psi_g}$, impose the equality and solve the integral in order to obtain $\tau\equiv T_{A}$. We stress that, following this procedure, the QSL time is defined as the time required by the process to traverse a distance equal to $s(\psi_0,\psi_g)$ in state space, regardless of the states being actually connected in the evolution \cite{bib:davidovich}. Another possibility, proposed in Ref. \cite{bib:lutz2}, is to replace the integral on the r.h.s. of eq. (\ref{ec:cota}) by the time-averaged energy variance
\begin{equation}
\overline{\Delta E(\tau)}=\frac{1}{\tau}\int_0^\tau\: \Delta E (t')\:dt',
\label{ec:vmedio}
\end{equation}
\noindent yielding
\begin{equation}
\tau\geq\frac{\mathrm{arccos}\left(\left|\BraKet{\psi_0}{\psi(\tau)}\right|\right)}{\overline{\Delta E(\tau)}}.
\label{ec:tlutz}
\end{equation} 
Evaluating this expression for $\tau=T$, we obtain
\begin{equation}
T\geq\frac{\mathrm{arccos}\left(\left|\BraKet{\psi_0}{\psi_g}\right|\right)}{\overline{\Delta E(\tau)}}\equiv T_{B},
\label{ec:lutz}
\end{equation}
\noindent which gives another version of the QSL time. In this case the geometrical interpretation is straightforward, as it is easy to check that $T_B=\frac{s}{s_p}\:T\leq T$, where $s$ is the distance between $\psi_0$ and $\psi_g$ measured by (\ref{ec:dist}), $s_p$ is the length of the actual path traversed by the system during the evolution and clearly $s_p\geq s$.\\


Finally, we mention that in some cases we could also obtain a lower bound on the time evolution from the geometrical relation by again evaluating for $\tau=T$, solving the integral in the r.h.s. and manipulating the result in order to reach an inequality of the form
\begin{equation}
T\geq T_C,
\end{equation}

this is, $T_C$ is the lower bound obtained by analytically working out the value of $T$ from the inequality (\ref{ec:cota}). \\

The three procedures we mentioned clearly give the same result for time-independent Hamiltonians, where the original MT bound is recovered. However, in the general time-dependent case they may differ, as we will show in the next section. Note that complete knowledge of the Hamiltonian at all times $H(t)$ is not enough to compute $\Delta E(t)$, since also the state of the system $\Ket{\psi(t)}$ is necessary.\\

\section{Discussion: driven two-level system} 

\begin{figure}[h]
\begin{center}
\includegraphics[width=0.6\linewidth]{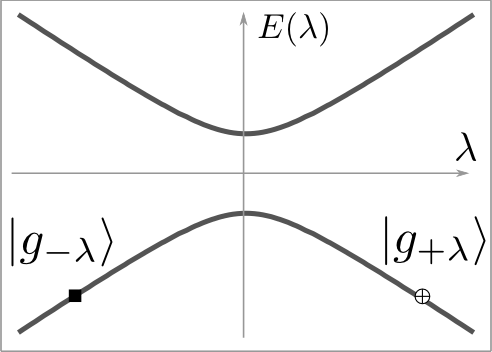}
\label{fig:fig1} 
\caption{Energy spectrum of Hamiltonian (\ref{ec:hami}) as a function of $\lambda$. Initial and final states for the control problem are shown with symbols $\blacksquare$ and $\oplus$.}
\end{center}
\end{figure}

\begin{figure}[h!]
\begin{center}
\includegraphics[width=0.85\linewidth]{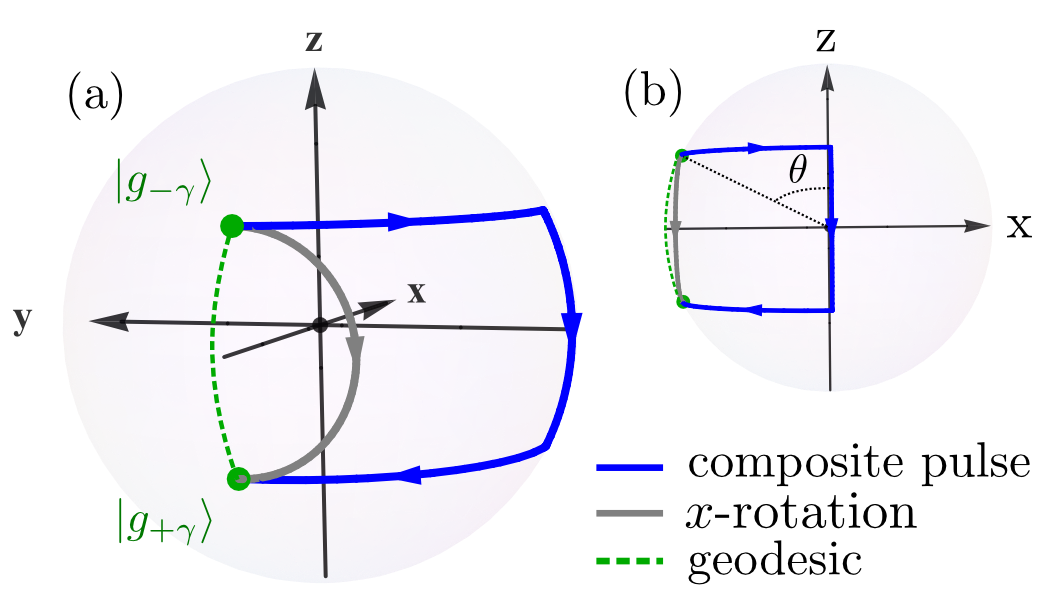}
\caption{\label{fig:fig2}(Color online) (a) Bloch sphere representation of the two-level system state space. Blue (dark) line shows the evolution generated by the composite pulse protocol, the gray line shows the evolution for $\lambda=0$ and the dashed line represents the geodesic path linking $\Ket{g_{-\gamma}}$ and $\Ket{g_{+\gamma}}$. We used $\omega=1$ and $\gamma=2$. (b) Same as (a) but from a different point of view.}
\end{center}
\end{figure}

We now show a few examples that illustrate the procedure for evaluating the QSL. Consider a two-level system with Hamiltonian
\begin{equation}
H(\lambda)=\omega\sigma_x+\lambda\sigma_z
\label{ec:hami}
\end{equation}
\noindent where $\omega$ is fixed, $\sigma_i$ denote the Pauli operators and $\lambda$ represents an external driving field. The energy levels of the system, as a function of $\lambda$, describe an  spectrum with an avoided crossing (AC) at $\lambda=0$, as can be seen in Fig. \ref{fig:fig1}. The energy eigenstates are function of $\lambda$ as well, and we denote them $\left\{\Ket{g_{\lambda}},\Ket{e_{\lambda}}\right\}$ for each $\lambda\in\mathbb{R}$, $g$ and $e$ representing ground (lower) and excited (higher) states, respectively. We pose the following control problem: consider the situation where the system is prepared at $t=0$ in $\Ket{\psi_0}=\Ket{g_{-\gamma}}$ and reaches $\Ket{\psi_f}=\Ket{g_{+\gamma}}$ at $t=\tau$, where $\gamma>0$. The distance between those states, measured by the distance defined in eq. (\ref{ec:dist}) is
\begin{equation}
s(\psi_0,\psi_f)=\pi-2\theta\equiv s(\theta),
\end{equation}
\noindent where we have defined $\mathrm{tan}(\theta)=\omega/\gamma$, see Fig. \ref{fig:fig2} (b).  If there are no constrains on the possible values of $\lambda$, the time-optimal control solution originally shown in \cite{bib:caneva} is the ``composite pulse protocol'', where $\lambda(t)$ takes the following form

\begin{equation}
\lambda(t)=\left\{
\begin{array}{l l}
\lambda_0 & \  0<t<t_0 \\
0 & \ t_0<t<T+t_0 \\
-\lambda_0 & \ T+t_0<t<T+2t_0
\end{array}
\right.,
\label{ec:opt}
\end{equation}

\noindent such that $\lambda_0\gg\omega$ and $\lambda_0 t_0=\pi/4$, in order to generate a $\pi/2$ rotation around the $z$-axis in the first and final step of the protocol. The middle step is a rotation around the $x$-axis. In Fig. \ref{fig:fig2} we show the overall evolution generated by this Hamiltonian in Bloch sphere. All shown trayectories were simulated by solving Schr\"odinger equation numerically (using the usual four-step Runge-Kutta method) with $\omega=1$, $\gamma=2$ and $\lambda_0=10$. \\

Note that $t_0$ can be chosen as close to zero as desired, so that the total duration of the protocol satisfies $\tau\equiv T+2t_0\cong T$. In a recent work \cite{bib:heger}, it was shown analytically that 
\begin{equation}
T=T(\theta)=\frac{1}{\omega}\mathrm{arctan}\left(\frac{\gamma}{\omega}\right)=\frac{s(\theta)}{2\omega}.
\label{ec:theger}
\end{equation}

In order to obtain the bounds $T_K$ (with $K=A,B,C$) discussed the previous section for this process, it is necessary to calculate the integral on the r.h.s. of expression (\ref{ec:cota}). To do so, we express the state of the system at time $t$ using the usual Bloch parametrization
\begin{equation}
\Ket{\psi}=\mathrm{cos}\left(\frac{\chi}{2}\right)\Ket{0}+\ex{i\varphi}\mathrm{sin}\left(\frac{\chi}{2}\right)\Ket{1},
\end{equation}
\noindent where $\chi=\chi(t)$ and $\varphi=\varphi(t)$ are the usual polar and azimuthal angles used in spherical coordinates. The variance of Hamiltonian (\ref{ec:hami}) can then be expressed as 

\begin{eqnarray}
\nonumber \Delta E^2&=&\lambda^2\:\mathrm{sin}^2\left(\chi\right)+\omega^2\left(1-\mathrm{sin}^2\left(\chi\right)\mathrm{cos}^2\left(\varphi\right)\right) \\ 
& &-2\:\lambda\:\omega\:\mathrm{sin}\left(\chi\right)\mathrm{cos}\left(\chi\right)\mathrm{cos}\left(\varphi\right). \label{ec:var}
\end{eqnarray} 

Due to the piecewise-constant time-dependance of $\lambda(t)$, i.e. expression (\ref{ec:opt}), this protocol has three steps. In the first one, $t\in\left[0,t_0\right]$ and the last one $t\in\left[T+t_0,T+2t_0\right]$,  $\chi=const.=\theta$ and $\varphi$ varies from $\pi$ to $\frac{3}{2}\pi$ (or in reverse) with angular velocity $2\lambda_0$. This results in
\begin{equation}
2\int_0^{t_0}\: \Delta E (t')\:dt'=2\int_{T+t_0}^{T+2t_0}\: \Delta E (t')\:dt'=\frac{\pi}{2}\:\mathrm{sin}\left(\theta\right),
\label{ec:primer}
\end{equation}
\noindent which, naturally, is the length of the path travelled by the system in each step. For $t\in\left[t_0,T+t_0\right]$, we have $\varphi=\frac{3}{2}\pi=const.$ and the polar angle runs from $\theta$ to $\pi-\theta$ with velocity $2\omega$, so we get

\begin{equation}
2\int_{t_0}^{t_0+T}\: \Delta E (t')\:dt'=2\int_{t_0}^{t_0+T}\:\omega\:dt'=2\omega T.
\label{ec:medio}
\end{equation}

As was discussed in the previous section, to obtain the bound $T_A$ we have to solve
\begin{equation}
s(\theta)=2\int_{0}^{T_A}\Delta E(t')\:dt',
\end{equation}

\noindent in which different results are be obtained depending on the values of $\theta$, yielding

\begin{equation}
T_A(\theta)=\left\{
\begin{array}{l l}
 0 & \mathrm{if} \ \frac{\pi}{2}\:\mathrm{sin}\left(\theta\right)\geq s(\theta) \\
\frac{s(\theta)-\frac{\pi}{2}\:\mathrm{sin}\left(\theta\right)}{2\omega} & \mathrm{if}\ \mathrm{not}\
\end{array}
\right.,
\label{ec:ta}
\end{equation}

\noindent where he have taken the limit $\lambda_0\rightarrow\infty$. Note that in this procedure the third step of the protocol does not need to be computed, since by that point the distance traversed by the system would surely be larger than $s(\theta)$. It can be seen by comparing  expression (\ref{ec:ta}) to eq. (\ref{ec:theger}) that $T_A<T$ for all $\theta$. The bound $T_B$, given by eq. (\ref{ec:lutz}), can also be evaluated directly and gives
\begin{equation}
T_B(\theta)=\frac{s(\theta)}{s_{path}(\theta)}T(\theta)=\frac{s(\theta)}{s(\theta)+\pi\mathrm{sin}\left(\theta\right)}T(\theta)
\end{equation}

Finally, for this particular process, is indeed possible to evaluate the integral and work out an inequality for the evolution time. This follows directly from replacing the results in eq. (\ref{ec:primer}) and eq. (\ref{ec:medio}) in the general expression (\ref{ec:cota})
\begin{equation}
s(\theta)\leq2\int_0^{T+2t_0}\Delta E(t')\:dt'=\pi\mathrm{sin}\left(\theta\right)+2\omega T,
\end{equation}
\noindent such that we get
\begin{equation}
T\geq\frac{s(\theta)-\pi\mathrm{sin}\left(\theta\right)}{2\omega}\equiv T_C(\theta).
\end{equation}
Note that $T_C$ can be negative for certain values of $\theta$, for which we will consider the higher bound $T_C=0$ as the physically meaningful one. \\

\begin{figure}[h!]
\begin{center}
\includegraphics[width=0.9\linewidth]{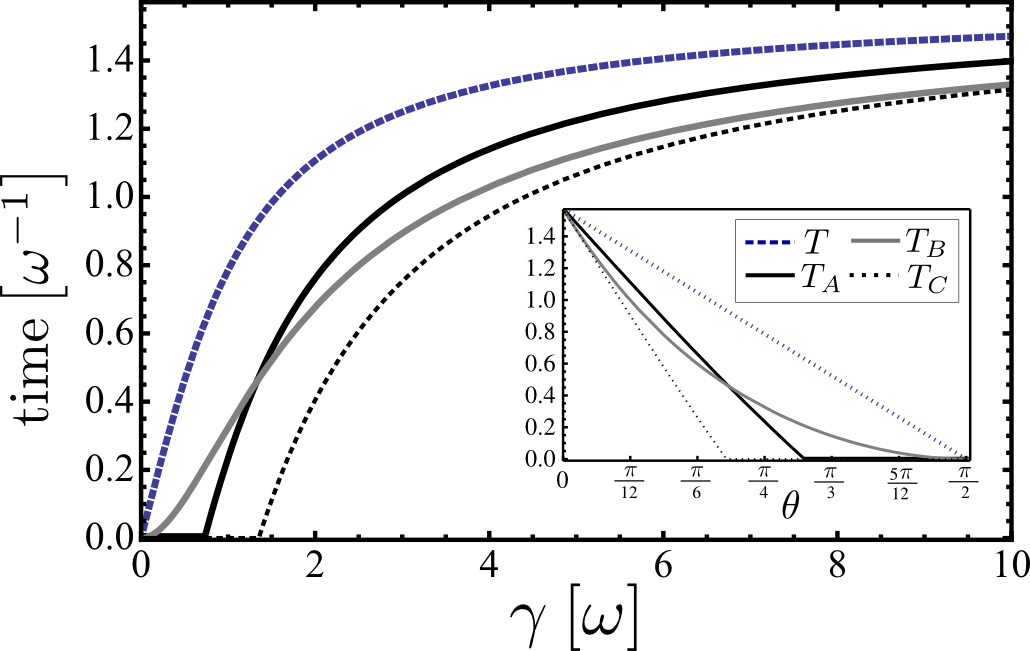}
\caption{\label{fig:plot1} (Color online) Optimal evolution time $T$ and bounds $T_A,\:T_B,\:T_C$ obtained from eq. (\ref{ec:cota}) for the composite-pulse protocol (with unconstrained $\lambda$) as a function of parameter $\gamma$. (Inset) Same plots as a function of $\theta=\mathrm{arctan}\left(\omega/\gamma\right)$, the azimuthal angle of the initial state in the Bloch sphere (see Fig. 2)}
\end{center}
\end{figure}

In Fig. \ref{fig:plot1} we plot the evolution time $T$ given by eq. (\ref{ec:theger}) along with the different bounds we have obtained, as a function of $\gamma$ for fixed $\omega$. In the inset of the figure, we show the same  as a function of $\theta$. Note that $\gamma$ determines the initial and final states of the process, and that when $\gamma=0$ ($\theta=\pi/2$), both states coincide since $\Ket{g_{\gamma=0}}=\Ket{\downarrow_x}$, i.e. the eigenstate of $\sigma_x$ with eigenvalue equal to $-1$. In the limit $\gamma\rightarrow\infty$ ($\theta\rightarrow 0$), the states tend to the orthogonal set $\left\{\Ket{\downarrow_z},\Ket{\uparrow_z}\right\}$, i.e., the eigenstates of $\sigma_z$. In the plot, it can be seen that in both limits the different bounds are equal and are saturated by the optimal time. This is trivial for $\gamma=0$ (for which $s(\theta)=0$), and it is also clear for $\gamma\rightarrow\infty$, since in this limit, the evolution is a rotation around the x-axis which connects the poles of the sphere through a geodesic. For finite $\gamma>0$, the evolution time $T$ is strictly higher than all three bounds, as expected, and $T_C$ is the lower bound. On the other hand, the plots of $T_A$ and $T_B$ cross for a certain value of $\gamma$, so that we cannot assert that one expression gives a tighter bound than another. Moreover, $T_A$ and $T_C$ vanish for a certain range of $\gamma$, meaning that in that regime, they do not give a meaningful limitation for the evolution time. Note that in the time-independent case, the MT bound (eq. \ref{ec:mtbound}) gives zero only if $\psi_0=\psi(\tau)$ (which is trivial) or if $\Delta E\rightarrow\infty$, which means that the system evolves uniformly with infinite velocity. In the time-dependent formulation, the velocity of the system in state space is not constant, and the QSL time can vanish if, for some period, $\Delta E\rightarrow\infty$. \\

\begin{figure}[h!]
\begin{center}
\includegraphics[width=0.5\linewidth]{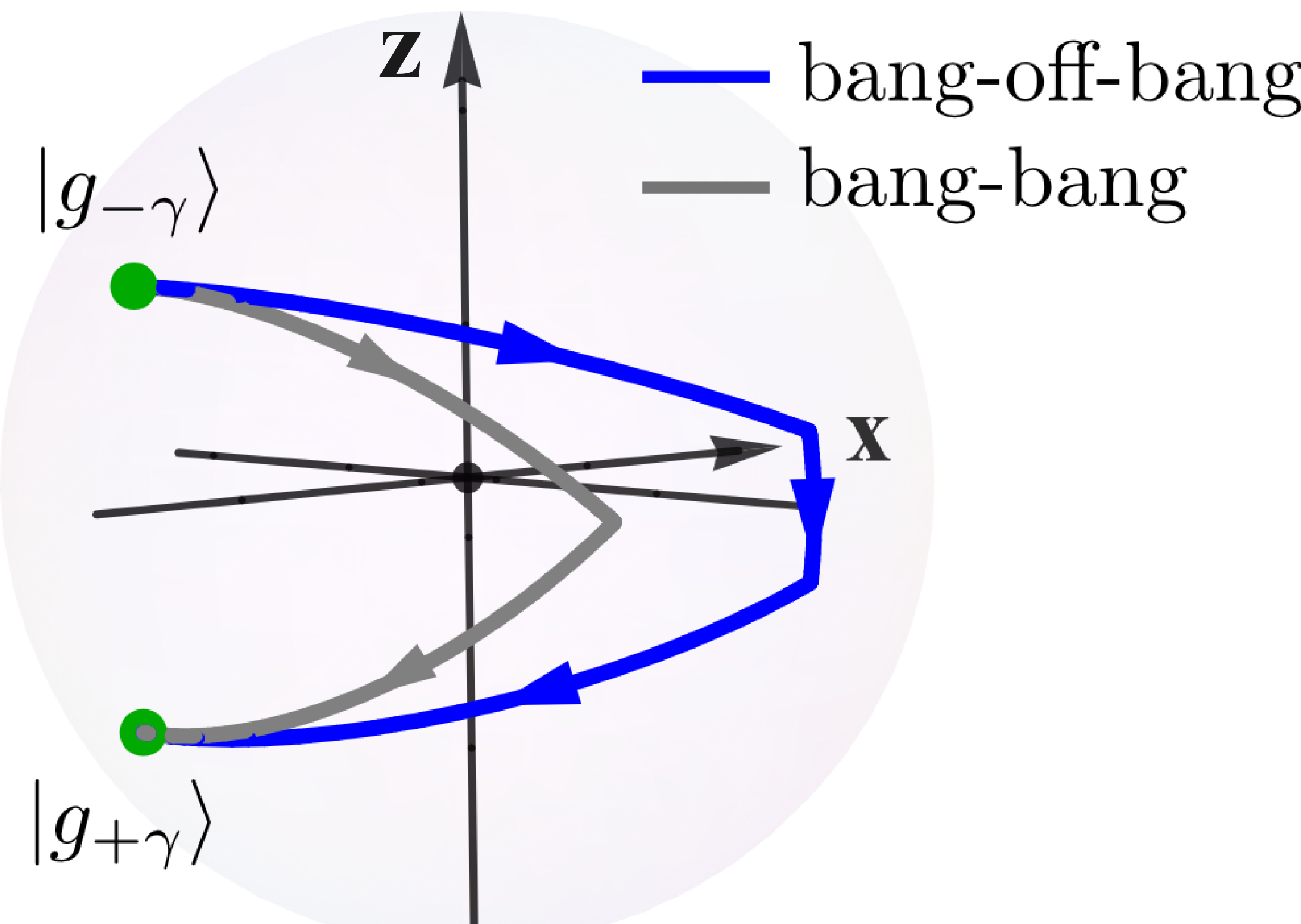}
\caption{\label{fig:fig3} Bloch sphere trajectories for the composite pulse protocol with constrained $\lambda$. Bang-off-bang protocol was simulated with $c=1.5\omega^2/\gamma$, while for the bang-bang protocol, $c=0.5\omega^2/\gamma$ was used. The values of $\omega$ and $\gamma$ used were the same as in Fig. \ref{fig:fig2}}
\end{center}
\end{figure}

\begin{figure}[h!]
\begin{center}
\includegraphics[width=0.8\linewidth]{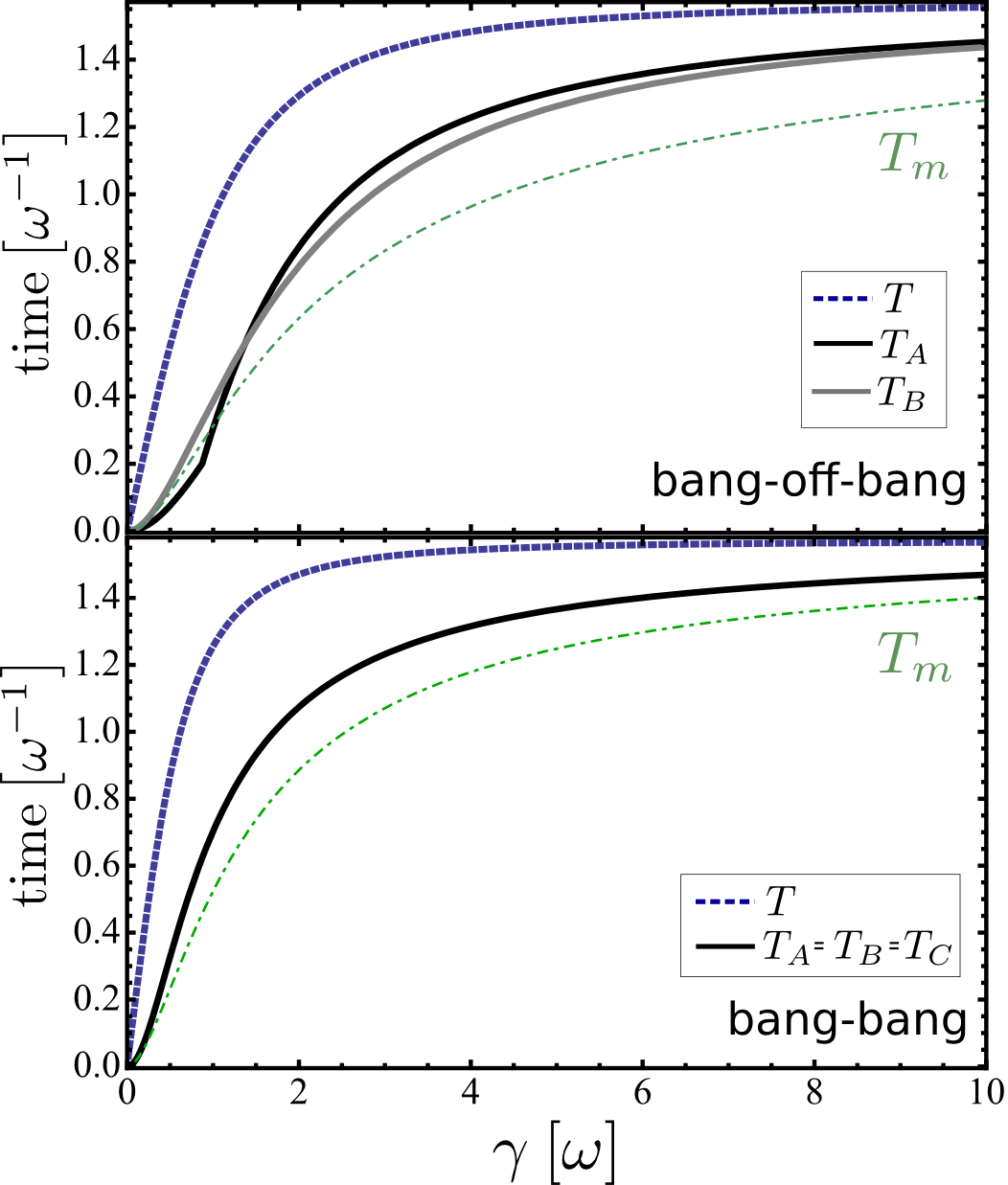}
\caption{\label{fig:plot2} (Color online) (top) Optimal evolution time $T$ and bounds $T_A,\:T_B$ obtained from eq. (\ref{ec:cota}) for the bang-off-bang protocol ($|\lambda_0|\leq c$, $c>\omega^2/\gamma$) as a function of parameter $\gamma$. (bottom) Same quantities for the bang-bang protocol ($|\lambda_0|\leq c$, $c<\omega^2/\gamma$). In both cases, the weaker bound $T_m$, which is discussed at the end of Section III, is displayed (dot-dashed line). See text for details.}
\end{center}
\end{figure}

In the previous example, the value of $\lambda(t)$ was unbounded, and so we could choose $\lambda_0\rightarrow\infty$ so as to generate instantaneous rotations around the $z$-axis. If the restriction $|\lambda_0|\leq c$ is added, the optimal solution (\ref{ec:opt}) changes, and different results are obtained wheter $c>\omega^2/\gamma$ or $c<\omega^2/\gamma$. In the first case, the optimal control protocol is of bang-off-bang type, meaning that the evolution is again in three-steps with $\lambda=0$ in the middle. In the latter, the protocol is of bang-bang type, so that $\lambda\neq0$ throughout the evolution. In both cases, $\left|\lambda(t)\right|$ takes its maximum possible value, that is, $c$. Detailed discussion about this cases can be found in Ref. \cite{bib:heger}. We show the trayectories generated by both protocols in Fig. \ref{fig:fig3}. We used $c=1.5\:\omega^2/\gamma$ for the bang-off-bang case and $c=0.5\:\omega^2/\gamma$ for the bang-bang protocol. Note that in both cases the initial and final rotations take place in a tilted axis in the x-z plane, and yield finite evolution time. The bounds described in the previous section can be obtained for these protocols (altough $T_C$ which cannot be worked out anallytically for the bang-off-bang case). We plot the optimal time along with these bounds in Fig. \ref{fig:plot2}. For the first protocol (top figure), we observe the same features as in the unconstrained case, i. e., the evolution time is strictly bounded by below as expected by $T_A$ and $T_C$ and all quantities are equal for $\gamma=0$ and $\gamma\rightarrow\infty$. Also, both bounds cross for certain $\gamma>0$. For the second protocol (bottom figure) all bounds yield the same result for every $\gamma$, due to the fact that in this particular case, $\Delta E$ is constant.\\

Having explored the bounds obtained directly from expression (\ref{ec:cota}) for the examples shown above, we remark that in all cases, considerable knowledge about the state of the system at all times was required to acquire those bounds. At the very least, both the total evolution time and the length of the path followed in state space is required (for obtaining $T_B$). For computing $T_A$ and $T_C$, we must know $\Delta E(t)$ at all times, which usually requires knowledge of the time evolution operator $U(t)$ for all $t\geq0$ or, at least, of $H(t)$ and $\Ket{\psi(t)}$. In the time-independent regime, given an initial and final state, only $\Delta E$ (which is constant) is required in order to evaluate the MT bound. So, in this regime, the QSL can sometimes be useful as a simple straighforward estimation for the minimum evolution time, which can be computed before analyzing the actual evolution of the system. For time-dependent systems, on the other hand, QSL times have to be obtained after the whole physical process is determined. \\

Of course, we can still obtain a lower bound on the evolution time which is computed in a simpler way. Namely, as $\Delta E(t)\geq0$ by definition on the integral on the r.h.s. of eq. (\ref{ec:cota}), it follows that
\begin{equation}
2\mathrm{arccos}\left(\left|\BraKet{\psi_0}{\psi(\tau)}\right|\right)\leq 2\int_0^\tau\: 2\Delta E (t')\:dt'\leq 2\Delta E^{max}\tau,
\end{equation}
\noindent where $\Delta E^{max}=\mathrm{max}_{0<t<\tau}\:\Delta E(t)$ and so
\begin{equation}
\tau\geq\frac{\mathrm{arccos}\left(\left|\BraKet{\psi_0}{\psi(\tau)}\right|\right)}{\Delta E^{max}}.
\label{ec:cotamax}
\end{equation}

This expression will in general be computable without knowing the complete form of $U(t)$ but will usually give a weaker bound for the evolution time. For the composite pulse protocol with unconstrained $\lambda$ discussed at the beginning of this section, it is clear that $\Delta E^{max}=\infty$, so that from relation (\ref{ec:cotamax}) we merely get $\tau\geq0$. Of course, for protocols which finite velocity of the system in state space, the bound will be greater than zero. This is the case of the bang-off-bang and bang-bang protocols, where $\Delta E$ can be bounded straightforwardly from eq. (\ref{ec:var})
\begin{equation}
\Delta E\leq\left|\lambda^{max}\right|+\left|\omega\right|=\left|c\right|+\left|\omega\right|.
\end{equation}

Then, expression (\ref{ec:cotamax}) yields
\begin{equation}
\tau\geq=\frac{\frac{1}{2}\left(\pi-2\theta\right)}{\left|c\right|+\left|\omega\right|}\equiv T_{m}.
\label{ec:tmax}
\end{equation}

In Fig. \ref{fig:plot2} we include the plot $T_m$ as a function of $\gamma$ for both brotocols (dot-dashed line), and it can be seen that they give weaker bounds on the evolution time than all the rest. \\

Finally, we remark that relation (\ref{ec:cota}) can give a tight bound in our examples if we follow a different procedure. Turning again to the unconstrained composite-pulse protocol, note that we know the state of the system as a function of time $\psi(t)$ given by this control protocol, and remember the Hamiltonian given by expressions (\ref{ec:hami}) and (\ref{ec:opt}) is piecewise-constant and consists on three steps. So, lower bounds $T_i^{min}$ on each step $i$ ($i=1,2,3$) of the procedure can be found by means of the time-independent MT bound (\ref{ec:mtbound}). The total evolution time $T$ then satisfies 
\begin{equation}
T=T_1+T_2+T_3\geq T_1^{min} + T_2^{min} + T_3^{min},
\label{ec:otrobound}
\end{equation}
\noindent where $T_i$ is the time required in step $i$ of the protocol. Evaluation of $T_i^{min}$ is straightforward from eq. (\ref{ec:mtbound}), but the initial and final states of each step has to be known. For this control protocol, clearly $T_1^{min}=T_3^{min}=0$ and
\begin{equation}
T_2^{min}=\frac{\pi-2\theta}{2\omega},
\end{equation}
\noindent so that expression (\ref{ec:otrobound}) yields
\begin{equation}
T\geq\frac{\pi-2\theta}{2\omega}.
\label{ec:thight}
\end{equation}
Comparing with the optimal result, eq. (\ref{ec:theger}), it can be readily seen that the equality in eq. (\ref{ec:thight}) holds and so the bound obtained is thight. Note that in step 2, the state follows a geodesic between the initial and final steps (see Fig. \ref{fig:fig2}).\\

\section{Conclusions}

In this paper we explored the results obtained from the usual Quantum Speed Limit formula (\ref{ec:cota}) for time-dependent systems when applied to a quantum control problem, for which optimal solutions are known. We show that a number of bounds on the evolution time can be obtained, which can be in general different for the same physical process. In our analysis, we discuss the specific meaning of the QSL time, which can be described as the minimum time required by a quantum system to traverse a certain distance in state space, under the action of a fully determined Hamiltonian. Also, we connect the QSL problem with quantum control, and point out that in some cases no meaningful bound for the total evolution time of a control protocol (i.e., only $T\geq0$) can be drawn from this expressions, a feature that is only possible in the time-dependent regime (in non-trivial cases). Finally, we remark that for time-dependent Hamiltonians, the QSL formulation in general requires knowledge about the state of the system at all times, but weaker bounds may be obtained by imposing restrictions on the parameters of the control Hamiltonian.

\acknowledgments

We would like to thank Ruynet Lima de Matos Filho for helpful remarks on the first version of the manuscript. We also thank Dorje Brody and Gonzalo Muga for useful discussions. We acknowledge the support from CONICET, UBACyT, and ANPCyT, Argentina.

\end{document}